\documentclass[twocolumn,prl,showpacs]{revtex4}
\usepackage{graphicx,epsfig,amssymb}

\begin{document}

\title{Preferential attachment in the protein network evolution}
 
\author{Eli Eisenberg and Erez Y. Levanon}
\affiliation{Compugen Ltd., 72 Pinchas Rosen Street, Tel Aviv 69512, Israel}

\begin{abstract}
The {\it Saccharomyces cerevisiae} protein-protein interaction map, 
as well as many natural 
and man-made networks, shares the scale-free topology. The preferential 
attachment model was suggested as a generic network evolution model that 
yields this universal topology. However, it is not clear that the model 
assumptions hold for the protein interaction network. 
Using a cross genome comparison we show that (a) the older a protein, 
the better connected it is, and (b) The number of interactions a protein 
gains during its evolution is proportional to its connectivity. Therefore,
preferential attachment governs the protein network evolution. 
Evolutionary mechanisms leading to such preference and some implications 
are discussed.
\end{abstract}

\pacs{89.75.Hc, 87.23.Kg, 89.75.Da}

\maketitle 

The analysis of networks has attracted great interest in recent years. 
Many man-made networks, including the World Wide Web\cite{1}, 
scientific\cite{2} 
and movie actor\cite{3} collaborations, and linguistic\cite{4} networks, 
have been shown to be scale free, with different nodes having 
widely different connectivities\cite{5,6,7}. Networks of biological 
origin, such as metabolic interaction\cite{8} and protein-protein 
interaction networks\cite{9}, also share this property. 
The emergence of the scale-free topology in such diverse examples calls for a 
universal explanation, based on generic principles, applicable to 
all the different networks studied. This was achieved by the growing 
network model, suggested by Barab\'asi and Albert\cite{10}, 
which assumes the continuous creation of new nodes and their
preferential attachment to previously well-connected nodes. 
An exact solution for the dynamics of the model demonstrates the 
emergence of the scale-free topology from 
these generic assumptions, given an asymptotically linear attachment 
kernel\cite{11, 12}. The model assumptions seem self-evident for 
social networks. A direct test for some 
of these networks have validated the preferential attachment principle, 
and shown an approximate linear kernel\cite{13, 14}. 
However, it is less clear how this model can be 
justified for natural networks, such as the biological networks.
While the dynamic growth of the network can be 
understood on an evolutionary time scale\cite{10}, 
the preferential attachment assumption is far from obvious,
as the interactions are not formed based on a conscious choice.

In this work, we focus on the {\it Saccharomyces cerevisiae} 
(bakers' yeast) protein-protein interaction network, 
which is often used as a model for a biological interaction network. 
A cross-genome comparison is employed to obtain a classification of the yeast 
proteins into different age groups. We observe a correlation between a 
protein's age and its network connectivity, in accordance with the growing 
network picture. Furthermore, this classification enables us to directly 
observe the preferential attachment phenomenon. Signs of this phenomenon
have been previously observed through  analysis of divergent pairs of
duplicated genes \cite{wagner}. 
We thus conclude that the Barab\'asi-Albert model is indeed relevant for 
describing the evolution of the yeast protein-protein interaction map. 
We further discuss implications of this phenomenon to the governing rules 
of protein evolution.

\begin{figure}[tb]
\includegraphics[width=2.4in]{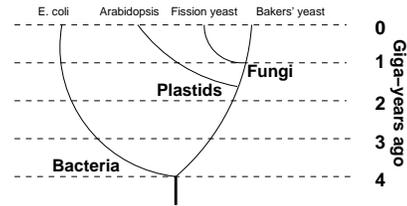}
\caption{A schematic representation of the relative position of the 
four studied organisms on the phylogenetic tree, based on Ref. \cite{16}.
The phylogenetic tree describes the evolutionary relationships between 
organisms. The root corresponds to the origin of life (first living cell),
and each branch point describes the emergence of distinct species out of 
one common ancestor. The evolutionary distance between any two organism is 
related to the sum of distances between each organism
and their closest common ancestor.
} 
\label{tree}
\end{figure}

We start by classifying the whole database of 6294 bakers' yeast 
proteins\cite{15} into four age groups. For this purpose, we pick
three other model organisms for which a fully sequenced genome and 
a comprehensive list of proteins are available, 
and are of varying evolutionary distance from the baker's yeast. 
The evolutionary distance between two organisms can be extracted from the 
phylogenetic tree (the "tree of life") describing the evolutionary 
branching process\cite{16} (see figure \ref{tree}): 
{\it Escherichia coli}\cite{17} belongs to the Bacteria branch 
(estimated time of diversion 4 Giga-years ago, Gya), 
{\it Arabidopsis thaliana}\cite{18} belongs the Plants branch 
(estimated diversion 1.6Gya), while {\it Schizosaccharomyces pombe}\cite{19} 
(fission yeast) and the bakers' yeast belong to different sub-phyla on the 
Fungi branch (estimated diversion 1.1Gya). 
A cross-genome comparison 
between these organisms is employed in order to estimate the age of 
each bakers' yeast protein. 
We assume that a protein created at a certain time in a certain ancestor 
organism will have descendants in all organisms that diverged from this 
ancestor. For example, proteins that are older than the first 
(Bacteria) diversion should have descendants in all four organisms, while 
those created after the fission-yeast diversion are expected to have 
descendants in the bakers' yeast alone. 
While the descendant proteins continue to evolve and diverge, 
they still show higher sequence similarity than a random pair of proteins.

For each of the bakers' yeast proteins, we search for similar proteins
in the other three organisms (see details below), 
and use the results to classify it into one of four age groups. 
Proteins with no fission-yeast similarities are expected to be relatively 
new (group 1, 872 proteins); those with similarities only in fission-yeast 
are expected to have an ancestor prior 
to the diversion and are therefore older (group 2, 665 proteins); those with 
fission-yeast and Arabidopsis similarities are even older (group 3, 2079 
proteins); and those with analogues in all three organisms form the 
oldest group of proteins (group 4, 2678 proteins), with ancestors that 
predate the first diversion. Only a small fraction (less than 10\%) of 
the similarities were not consistent with the evolutionary timeline. 
Note that our age-group classification is not sensitive to duplication events 
\cite{duplications}, and thus new proteins generated by duplication are here 
classified as old. 

Here are some brief technical details on the similarity search done.
We use the standard definitions for the similarity distance between 
sequences, and employ the standard Protein-BLAST program\cite{20}. 
The program is given a query sequence (in our case: the yeast protein) 
and a reference database (the set of all proteins of the other organism), 
and compares the query
sequence to each of the database sequences, in search for shared patterns. 
Each found match gets a score (termed ``E-score''), which is the expected 
number of same or higher quality matches given a randomized database. 
The probability to get a match of same or higher quality for a random 
sequence is $$P(E|{\rm random\ pair})=1-\exp(-E),$$ where $E$ is the E-score. 
The lower this probability, the higher the confidence that the 
sequences similarity (or the match) is indeed due to a common ancestor 
for both sequences. We considered two proteins to be similar if the 
E-score of their match was lower than the cutoff value $E_c=0.7$, 
corresponding to $P(E_c|{\rm random\ pair})\simeq 0.5$.

\begin{figure}[tb]
\includegraphics[width=2.4in]{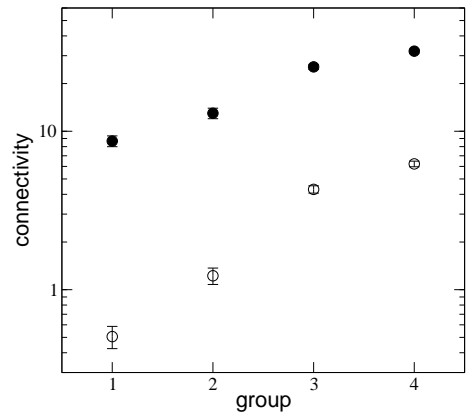}
\caption{Connectivity dependence on protein age. Averaged connectivity for 
four age 
groups of yeast proteins. Groups are numbered in increasing age order: group 1 
proteins (those with no similarities in fission-yeast, Arabidopsis or E.coli 
genomes) 
are expected to be the newest, and group 4 proteins (with similarities in all 
three 
organisms) are expected to be the oldest. Results are presented for the whole 
interactions database (solid symbols), and for a restricted set excluding the 
low-confidence interactions (open symbols). For most data points, the 
error-bar is smaller than the symbol}
\label{age}
\end{figure}

In the following, we use the obtained age-group classification of the 
yeast proteins to analyze the structure of the protein-protein interaction 
network. 
We use a published database of yeast protein-protein interactions\cite{21}, 
and first look at the average connectivity. Figure \ref{age} 
shows a clear dependency of the connectivity on the protein age, with older 
proteins having significantly more interactions. While group 1 proteins 
(newest) have only 0.5 links per protein, group 4 proteins (oldest) have 
6.2 links per protein. This supports the picture of the growing network 
model, where the older a node the 
higher its probability to gather interactions with other late-coming proteins. 

\begin{figure}[tb]
\includegraphics[width=3.2in]{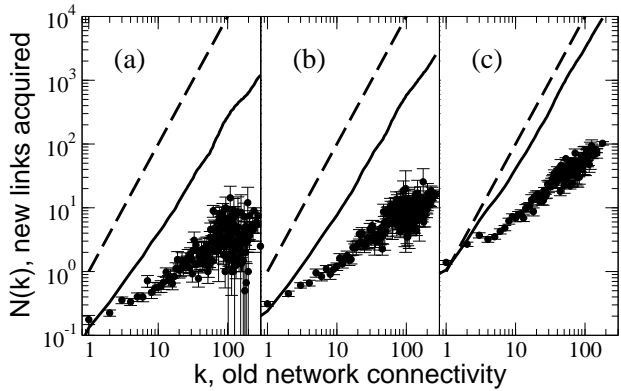}
\caption{Preferential attachment in protein network evolution. Symbols: The 
averaged 
number of links a protein acquires to proteins from new groups $N(k)$, 
as a function of $k$, its number of connections to all other (older) proteins. 
In order to study the asymptotic behaviour and estimate the exponent,
we plotted (solid lines) the integrated function 
$\kappa(k)\equiv\int_0^k N(x)dx$. 
An asymptotic power-law scaling $\kappa(k)\propto k^{\alpha+1}$ is observed 
with $\alpha\approx 1$, suggesting a linear preferential attachment 
kernel. The dashed line describes the power law function $k^2$, and is 
presented for 
comparison. Results have been obtained using the full interactions 
database\cite{21}.
(a) new links to proteins from group 1 alone, 
as a function of the number of links in groups 2, 3 and 4. 
(b) new links to groups 1 and 2. (c) new links to groups 1, 2 and 3 for all 
group-4 proteins. }

\label{pa}
\end{figure}

A direct test of the second assumption of the growing network model, namely, 
the preferential attachment principle, requires detailed information on the 
network development, which is beyond our reach. However, the above 
classification provides 
us with snapshots of the growing network at three points in its evolution, 
enabling an insight into the evolution of protein interactions.  
We study the sub-network 
defined by group 4 proteins and the links connecting them, recording the 
connectivity of each old protein on this sub-network. This sub-network was 
used as a model for the interaction map at an early stage of the evolution 
process (the time of divergence of the Bactria branch). The 
number of links of each old protein to the newer proteins (groups 1,2,3) is 
the number of links acquired since that time. We then looked at the number 
of new links a node gathered as a function of its connectivity in the old 
network. A similar analysis is done for the sub-networks defined by groups 
3 and 4 combined (proteins with an Arabidopsis analogue), and for groups 
2,3 and 4 combined (proteins with fission-yeast analogue). As Figure 
\ref{pa} shows, the number of new links tends to increase 
with the number of links in the old network, which is a signature of 
preferential attachment. The number of new links appears to be 
approximately linear in the connectivity, suggesting a linear 
preferential attachment kernel, and consistent with the scale free 
topology\cite{11}. 

The growing network paradigm suggests a dynamic model for preferential 
attachment: that is, all nodes are created equal and the attachment 
probability is related to the actual current connectivity ("rich get 
richer" model) as defined by the network dynamics. An alternative 
model\cite{22} suggests a static explanation in which each node 
has a different intrinsic fitness that determines its ability to interact and 
doesn't change as the network grows. In this model both the actual 
connectivity and the attachment probability of a protein depend on 
its intrinsic fitness. Given an appropriate distribution of the fitness 
parameter, this model can explain the results of figure \ref{pa} 
("good get richer" model), but it is not consistent with the age-dependence 
shown in figure \ref{age}. While the growing network model predicts 
that older nodes will be better connected, connectivity in the static model 
is related solely to the node fitness, and age and connectivity shouldn't 
be correlated. Thus, our results (figure \ref{age}) 
support the first option as a model for the protein interaction 
evolution. Gene duplication was also suggested as an explanation for the 
scale-free topology of the protein interaction network\cite{23,24,25,26}. 
However, since duplication events are not detected by our age-group 
classification, our results show that the proteins network structure 
cannot be attributed solely to evolution by duplication.

The question of the evolutionary mechanism leading to the dynamic preferential 
attachment remains: how does becoming better connected make a protein more 
attractive for future interactions, and why is the preference linear in 
the number of connections? 
We suggest two possible mechanisms that partially answer these questions: 

(i) The more connections a node acquires, the stronger is the selective 
pressure to make it more connectable. On the molecular level this can be 
understood as a tendency to increase the number of protein attachment 
domains\cite{domains} (such as the WW\cite{WW} or proline-rich\cite{proline} 
domains), or to improve the existing domains such that they bind to more 
target proteins. In this mechanism the preferential attachment is related
to the physico-chemical properties of the highly-connected protein.
In order to test this possibility, one can look at the
distribution of domains and other reoccuring patterns in the set of 
highly-connected proteins, and check whether connectability can be traced
to sequence motifs. However, the lack of well-studied interaction network 
for other organisms and the partial understanding of attachmnet properties 
of protein domains limits our ability to perform such study.

(ii) Many protein interactions are actually physical interactions that 
change or regulate the functionality of the interacting 
parties, such as phosphorylation and complex formation. 
The number of potential distinct operation modes of a protein
increases exponentially with the number of its regulating proteins,
and similarly the number of potential variants of a given complex
increases exponentially with the number of its building-block proteins.
Therefore, the more connected a protein, the stronger 
the selection towards creating a protein to interact with it. Here, the 
phenomena relates to the biological functionality of the protein.
This mechanism can be validated by the following experiment: 
current technology enables us to dig out proteins that form a complex
together with a given target protein\cite{tap}. One can look at 
the different complexes generated under varying conditions and study the 
different combinations obtained, that is, how many distinct complexes
were formed using the target protein. Then, it is possible to study
how many new structures have been made available by each complex member.
We predict that the contribution of each new member will be multiplicative,
i.e., the number of new structures will be, on average, proportional to
the total number of structures.

The preferential attachment phenomenon demonstrates an important principle in 
the process of evolution. It dynamically leads to the formation of big protein 
complexes and pathways, which introduce high complexity regulation and 
functionality. New systems are not generated as self-interacting modules of 
new proteins; rather, new proteins tend to connect to the old 
well-connected hubs of the network and modify existing functional units. 
Indeed, 267 of the 872 
group 1 proteins (31\%, versus 12\% of group 4) have no interactions 
documented in the database, indicating a very low number of actual 
interactions. Thus, we get information on protein's centrality based on its 
sequence alone. This information is helpful in analyzing the protein 
interaction network given the partial information available.

It was shown that the higher the connectivity of a node, the higher its 
probability to be essential, i.e. to have a lethal knockout phenotype\cite{9}. 
As mentioned above, highly connected nodes tend to be older. We find that 
essential proteinss also tend to be older: 
only 8\% of the newest proteins are essential, 
in contrast to 20\% of the oldest 
proteins ($\chi^2$-test p-value $3\cdot 10^{-20}$).  

In conclusion, we show that the protein networks evolve by creating new, 
unconnected links, which attach to the existing network according to the 
linear preferential attachment principle. This explains the scale-free 
topology shared by the network, and has implications for understanding the 
evolutionary mechanisms. The correlation 
of the protein's age to its centrality opens new possibilities for deriving 
information on the interaction network topology based on sequence data.

\begin{acknowledgments}
We thank A.-L Barab\'asi, Y. Kliger, A. Lipshtat and S. Redner for valuable 
discussions and comments. 
\end{acknowledgments}


\begin{thebibliography}{10}
\bibitem{domains}Domains are functional units of proteins, each domain 
having its unique role. The WW domain\cite{WW} and the proline-rich domain 
\cite{proline} are of the most versatile protein-protein interaction modules. 
Proteins with such domain will potentially have many interactions.
\end{thebibliography}
\end{document}